\documentclass[aps,draft,12pt,eqsecnum,amsmath,amssymb,floatfix]{revtex4}
\usepackage{epsfig}
\input epsf
\newcommand{\beq}{\begin{equation}}
\newcommand{\beql}[1]{\begin{equation}\label{eq:#1}}
\newcommand{\eeq}{\end{equation}} 
\newcommand{\beqn}{\begin{eqnarray}}
\newcommand{\eeqn}{\end{eqnarray}}

\newcommand{\tr}{{\rm tr}}
\newcommand{\pa}{\partial}

\newcommand{\csA}{{a}}
\newcommand{\csB}{{b}}

\newcommand{\Bf}{B_{\rm eff}}

\newcommand{\myfig}[3]{\begin{figure}[ht]
\begin{center}
\leavevmode
\epsfxsize=#2cm
\epsfbox{#1}
\end{center}
\caption{#3}
\label{fig:#1}
\end{figure}}

\begin{document}
\title{Non-commutative Chern-Simons for\\ the Quantum Hall System and Duality}
\author{Eduardo Fradkin, Vishnu Jejjala and Robert G. Leigh}
\affiliation{
   	Department of Physics,
    	University of Illinois at Urbana-Champaign,
       1110 W. Green Street,
    	Urbana, IL 61801,
    	E-mail: \email{rgleigh@uiuc.edu,efradkin@uiuc.edu}}
\date{\today}
\begin{abstract}
The quantum Hall system is known to have two mutually dual Chern-Simons
descriptions, one associated with the hydrodynamics of the electron
fluid, and another associated with the statistics. Recently, Susskind
has made the claim that the hydrodynamic Chern-Simons theory should be
considered to have a non-commutative gauge symmetry. The statistical
Chern-Simons theory has a perturbative momentum expansion. In this
paper, we study this perturbation theory and show that the effective
action, although commutative at leading order, is non-commutative. This
conclusion is arrived at through a careful study of the three-point
function of Chern-Simons gauge fields. The non-commutative gauge
symmetry of this system is thus a quantum symmetry, which we show can
only be fully realized only through the inclusion of all orders in
perturbation theory. We discuss the duality between the two
non-commutative descriptions.
\end{abstract}

\preprint{ILL-(TH)-02-xx}
\maketitle


\section{Introduction}

Fractional quantum Hall (FQH) states are topological quantum fluid
phases of two-dimensional electron gases in large magnetic fields. These
incompressible fluid states have universal wave functions, the Laughlin
wave functions~\cite{laughlin} and their hierarchical
extensions~\cite{haldane-hierarchy,halperin-hierarchy,jain}, and their
associated spectrum of low-energy excitations have universal quantum
numbers, such as their charge and statistics. Smooth short-distance
changes in the precise form of the electron-electron interactions do not
change the universal properties of these ground states and of their low
energy spectra. On closed surfaces, the ground states of FQH systems
exhibit degeneracies which are a direct manifestation of the topological
nature of these fluid states~\cite{wen-niu,wen-topo}.

The universal properties of FQH states can also be described in the form
of an effective quantum field theory for the degrees of freedom with
energies low compared to the cyclotron energy $\hbar \omega_c$ and/or
the Coulomb energy $e^2/(\epsilon \ell)$ (whichever is smallest), and at
distances long compared to the magnetic length $\ell=\sqrt{\hbar c/eB}$
(and/or any other  short distance scale related to the interaction). The
low energy theory of the FQH states is a Chern-Simons gauge theory.
There are two equivalent, mutually dual, descriptions of this effective
theory. In the first, the Chern-Simons gauge theory is derived directly
from a theory of interacting
electrons~\cite{zhk,lopez-fradkin-91,zhang}, and the gauge field is the
statistical gauge field. The second description is based on a
hydrodynamic
picture~\cite{blok-wen,wen-review,wen-zee-matrix,frohlich-zee,frohlich-kerler,
bahcall-susskind}. In this picture the gauge fields embody the conserved
charge densities and currents of the electron fluid. Naturally, both
descriptions, which are dual to each other, yield an effective action
for slowly varying electromagnetic gauge fields which also has a
Chern-Simons form, a result required by gauge invariance and broken
time-reversal symmetry. Thus Chern-Simons gauge field theories appear as
the natural field-theoretic description of the low energy physics of FQH
states.

In Ref. \cite{Suss}, Susskind argued that quantum Hall systems are
inherently non-commutative, and that their effective field theory should
be a non-commutative extension of the Chern-Simons gauge theory. From a
heuristic point of view, this is expected, because of the
non-\-commutative magnetic algebra of the quantum mechanics of electrons
in a magnetic field~\cite{girvin-jach}. Thus, when acting on the Hilbert
space of excitations of an incompressible (gapped) state, physical
observables such as the charge and current densities, obey a local
extension of the magnetic algebra~\cite{girvin-jach}, a magnetic current
algebra related to the $W_\infty$ symmetry of area preserving
diffeomorphisms~\cite{sakita,stone,ctz}. The physical observables of
this magnetic current algebra obey Moyal products controlled by the
magnetic length $\ell$.

The non-commutative gauge theory that Susskind discusses is an effective
hydrodynamic theory of the fractional quantum Hall states. Thus, as such
it describes only the low energy physics of these incompressible fluid
states. It is an extension of the conventional effective Chern-Simons
descriptions which have been used quiet successfully  to describe
interesting phenomena such as edge states. Susskind's non-commutative
hydrodynamics  is a natural extension of these theories. However,
Susskind's proposal poses a number of questions. For instance, the
non-commutativity of Susskind's theory, at the level of the effective
action of Ref.\ \cite{Suss}, is controlled not by the magnetic length
$\ell$, but instead by the particle density $\rho$ (or rather, by the
mean separation between the electrons). It is thus natural to inquire
what is the connection between Susskind's non-commutative Chern-Simons
gauge theory and the magnetic current algebra as they superficially
appear to be controlled by different (although obviously related)
non-commutativity parameters. On the other hand, since the
non-commutativity happens on such a short-distance scale, it is also
necessary to explain in what sense this theory is robust, {\it i.e.,}
insensitive to other non-universal short distance physical phenomena not
considered in this theory. For instance, at short distances the
particular form of the interaction matters since it is what determines
how big the energy gaps are and how fast do the excitations move. We
will see below that what is special about the non-commutativity is that
it is the only short distance effect associated with the broken
time-reversal symmetry and parity of electrons in magnetic fields, and
that in this sense this structure is robust. Nevertheless,  the effects
of non-commutativity show up primarily in non-uniform states and in
practice both the time-reversal even and odd processes are important.
For example, it is well known from the classic work of  Girvin,
MacDonald and Platzman~\cite{gmp} that the spectrum of collective modes
with a finite length scale, such as the roton modes, is due to a
combination of the magnetic algebra and electron-electron interaction
effects.

In this paper, we provide a complementary derivation of Susskind's
result, based directly on the field theory of interacting electrons
coupled to gauge fields. Our calculation is an extension of the standard
methods of Ref.\ \cite{lopez-fradkin-91}, which were used before to
derive the conventional Chern-Simons effective gauge theory and the
universal physics they contain: the values of the quantum Hall
conductivity for which the FQH state is stable, the ground state
degeneracy, and the quantum numbers of the excitations. Here we will
show that a non-commutative Chern-Simons gauge theory emerges naturally
within this framework. We will find that to lowest order of
approximation, the non-commutativity parameter does not agree  with the
one found by Susskind. However,  we will show that Susskind's result is 
required by the constraints imposed by the magnetic current algebra of
the charge densities and currents. Hence, higher order quantum
corrections lead to finite renormalizations of the effective
non-commutativity parameter. In contrast the level of the Chern-Simons
theory is not renormalized beyond its Gaussian value, due to the
constraints of topological and gauge invariance, and incompressibility.
We will show that these results are universal in the sense that these
are the only effects at this length scale which are odd under
time-reversal (and parity).

Throughout the paper, we shall use the following notation:
\begin{eqnarray}
a_\mu: && {\rm statistical\ CS\ gauge\ field}\nonumber\\
b_\mu: && {\rm hydrodynamic\ CS\ gauge\ field}\nonumber\\
A_\mu: && {\rm electromagnetic\ gauge\ field}\nonumber
\end{eqnarray}
The paper is organized as follows. In Section \ref{sec:hydro}, we
briefly review Susskind's fluid dynamics derivation of non-commutative
Chern-Simons theory for the Laughlin FQH states. In Section
\ref{sec:cs}, we review the derivation of commutative Abelian
Chern-Simons theory for the FQH states in the Jain sequences, based on
the notion of flux attachment, as the effective field theory of FQH
states. In Section \ref{sec:calc}, we present the results of our
calculation of the effective gauge field action, obtained by integrating
out fermions. The bulk of this calculation, as well as notational
details are relegated to Appendix \ref{sec:appendix}. In Section
\ref{sec:ncdual}, we discuss the non-commutative version of Chern-Simons
duality. Here we discuss the connection between non-commutative
Chern-Simons theory, the underlying magnetic algebra and the $W_\infty$
algebra of the currents. Finally, in Section \ref{sec:conclusions} we
present our conclusions.

\section{Hydrodynamic Chern-Simons Theory and Non-Commutativity}
\label{sec:hydro}

Susskind's discussion concerns the hydrodynamic Chern-Simons field. We
briefly review that derivation here, largely to establish notation. In
the hydrodynamic limit, we have a charged fluid of number density $\rho$
in a magnetic field $B$. A fluid element is labeled by coordinates
$x^a(y)$, the $y^i$ being co-moving coordinates. The Lagrangian is taken
to be
\begin{equation}
{\cal L} = \int d^2y\ \rho_0 \left( \frac{M}{2} \dot x_a^2 - 
\frac{eB}{2} \epsilon_{ab} x^a \dot x^b - V(\rho) \right).
\end{equation}
The gauge generators are of the form
\begin{eqnarray}
G & = & \frac{1}{M} \int d^2y\, \pi_a \delta x^a \\ \nonumber
  & = & \rho_0 \int d^2y\, \Lambda(y) \frac{\partial}{\partial y^i} \left[ \epsilon^{ij} 
        \left( \dot x_a + \frac{\omega_c}{2} \epsilon_{ab} x^b \right) 
	\frac{\partial x^a}{\partial y^j} \right] 
	\nonumber
\end{eqnarray}
where $\zeta^i(y) = \epsilon^{ij} \partial_j \Lambda$ is the
transformation of $y^i$. This symmetry, when restricted to the lowest
Landau level, reduces to area preserving diffeomorphisms (APD) in space.
The conservation of $G$ implies that
\begin{equation}\label{eq:consstuff}
V_\Sigma + \frac{e}{M} \Phi_\Sigma = {\rm constant}
\end{equation}
where $V_\Sigma$ is the vorticity and $\Phi_\Sigma$ is the magnetic flux
through a region $\Sigma$.

Assuming that the fluid is near static equilibrium at
$\rho\simeq\rho_0$, we may expand
\begin{equation}
x^a=y^a+\theta\epsilon^{ab} \csB_b(y)
\end{equation}
where $\theta\equiv 1/2\pi\rho_0$. The Lagrangian becomes
\begin{equation}
{\cal L} = \frac{M\theta}{4\pi} \int d^2y\, 
\left[ \omega_c \epsilon^{ij} \dot \csB_i \csB_j+ 
\dot \csB_i^2   -   c_s^2 (\nabla \times \csB)^2 \right]
\end{equation}
while eq. (\ref{eq:consstuff}) becomes
\begin{equation}
\frac{e}{M} \Phi_0 +\omega_c\theta \oint \csB - 
\theta^2 \oint d\csB_i \left( \dot \csB^i + 
\frac{\omega_c}{2} \epsilon^{ij} \csB_j \right)
= \rm{constant}
\end{equation}
In the vacua of interest, the vorticity vanishes, and we find the
constraint
\begin{equation}
\epsilon^{ij} \frac{\partial}{\partial y^j} \left[ \omega_c\csB_i - \theta\left( 
\dot \csB^k + \frac{\omega_c}{2} \epsilon^{kl} \csB_l \right) 
\frac{\partial \csB_k}{\partial y^i} \right] = 0
\end{equation}
This may be enforced by introducing a Lagrange multiplier $\csB_0$, and
consequently, we find that the action is (at leading order in powers of $B$)
\begin{equation}
S=\frac{\hbar\theta}{4\pi\ell^2}\int d^3y\ \epsilon^{\mu\nu\rho}
\left[ \csB_\mu\partial_\nu \csB_\rho
+\frac{\theta}{3}\epsilon^{ij}\partial_i \csB_\mu \partial_j \csB_\nu \csB_\rho\right]
\end{equation}
Now the supposition is that this action should be interpreted as the
leading term in a momentum expansion of
\begin{equation}
S_{NCCS} =\hbar \frac{k}{4\pi} \int d^3y\, \epsilon^{\mu\nu\lambda} 
\left( \csB_\mu \star \partial_\nu \csB_\lambda - 
\frac{2i}{3} \csB_\mu \star \csB_\nu \star \csB_\lambda \right)
\label{eq:NCCS}
\end{equation}
where the $\star$-product is
\begin{equation}
f(x) \star g(x) = \left. \exp\left\{\frac{i}{2} \theta \epsilon^{ij} 
\frac{\partial}{\partial y^i} \frac{\partial}{\partial z^j}\right\}
f(y) g(z) \right|_{y=z=x}
\label{eq:Moyal}
\end{equation}
Thus the field $b_\mu$ is a (hydrodynamic) non-commutative Abelian
Chern-Simons gauge field. The non-commutativity  is controlled here by
the parameter $\theta=1/2\pi\rho_0=\ell^2/\nu$. If we couple
this to the electromagnetic potential and compute the Hall conductance,
we find that the filling fraction is given by $\nu=1/k$. Hence, this
theory describes the quantum hydrodynamics of fractional quantum Hall
states in the Laughlin sequence $\nu=1/k$, with $k$ an odd integer.

\section{Duality Invariant Formalism: Commutative Limit} \label{sec:cs}

Now let us review the dual Chern-Simons pictures of the FQH states
following the arguments of Ref.\ \cite{chetan} and Ref.\
\cite{lopez-fradkin-98}. Let us describe the problem of electrons in a
large magnetic field in a Feynman path-integral picture. We will denote
the histories of the electrons in $2+1$-dimensional space-time by a set
of conserved $3$-currents $j_\mu$ (with $\mu=0,1,2$). Let us write the
(fermionic) path-integral in the form of a sum over current
configurations ({\it i.\ e.\/} a sum over the histories of the
particles)
\begin{equation}
Z=\sum_{[j]} e^{\displaystyle{iS[j]/\hbar-i\frac{e}{\hbar c}\int d^3x\ A_\mu j^\mu+i\phi[j]}}
\label{eq:Zf}
\end{equation}
where $\phi[j]$ is a phase which accounts for the statistics of the
particles. Here, $S[j]$ is the action for $N$ interacting fermions,
$A_\mu$ is the external magnetic field.

Next, we note that for every configuration of currents $[j]$ it is
possible to compute the linking number $\nu_L([j])$ of the
configuration, which is a topological invariant and it is an integer.
(Technically this requires the assumption that the currents do not cross
or equivalently that there is a ``hard core" condition.) The expression
of $\nu_L[j]$ in terms of $j$ is known as the Gauss invariant, which is
non-local:
\begin{equation}
\nu_L[j]=\int\int j\wedge \partial^{-1} \cdot j
\label{eq:link}
\end{equation}
We will express the fact that the currents $j_\mu$ are conserved,
{\it i.e.} $\partial \cdot j=0$ by introducing~\cite{wen-review} a
hydrodynamic field $b$, related to the particle current by
\begin{equation}
j^\mu=\frac{1}{2\pi} \epsilon^{\mu \nu \lambda} \partial_\nu b_\lambda
\label{eq:j}
\end{equation}
which we will denote in short hand form as $j=\frac{1}{2\pi} \partial \wedge b$.

When written in terms of $b$, the link invariant is
local and it has a Chern-Simons form~\cite{wilczek-zee,wu-zee}:
\begin{equation}
\nu_L[j]=
\frac{1}{4\pi^2}\int d^3 x \; b \cdot \partial \wedge b
\label{eq:linkb}
\end{equation}
Hence, for any arbitrary integer $n$, we can shift the action $S[j] \to
S[j]+2\pi i n \hbar\nu_L[j]$ without changing any quantum mechanical
amplitude.

To enforce current conservation, we insert $b_\mu$ into the partition function by making use of the identity 
\begin{equation}
1=\int [Db] \; \delta\left(j-\frac{1}{2\pi} \partial \wedge b \right)=
\int [Db][Da] \; \; e^{\displaystyle{i\int\; d^3x\;  
a \cdot \left(j-\frac{1}{2\pi} \partial \wedge b\right)}}
\label{eq:one}
\end{equation}
To represent the $\delta$-function, we have introduced another vector
field $a$. Thus, we may write the full fermionic path integral in the
equivalent form
\begin{equation}
Z=\int [Db][Da]\sum_{[j]} e^{\displaystyle{iS[j]/\hbar+i\phi[j]+
i\int d^3 x 
\left((a-\frac{eA}{\hbar c}) \cdot j-\frac{1}{2\pi}  a\cdot  \partial \wedge b+
\frac{2n}{4\pi}b\cdot \partial \wedge b\right)}}
\label{eq:Z-final}
\end{equation}
Hence, a system of interacting fermions in an external electromagnetic
field is equivalent to another system  with two gauge fields, $a_\mu$
and $b_\mu$, with the fermions minimally coupled to $a_\mu-(e/\hbar c)
A_\mu$. The gauge field $a_\mu$ is the statistical gauge field of Refs.\
\cite{zhk,lopez-fradkin-91,hlr}, and it corresponds to $2n$ fluxes being
attached to each fermion. Conversely, the gauge field $b_\mu$ is the
hydrodynamic field used by Wen~\cite{wen-niu,wen-review,
wen-zee-matrix}. As noted in Ref.\ \cite{lopez-fradkin-98}, the
effective action of  Eq.\ (\ref{eq:Z-final}) obeys the Chern-Simons
level quantization condition~\cite{witten-cs,hosotani}, and thus it is a
consistent form of flux attachment, on both open and closed manifolds.


\section{The Non-Commutative Effective Action for the Statistical Chern-Simons Field}
\label{sec:calc}

We will now use this description of interacting electrons in magnetic
fields to give a derivation of Susskind's non-commutative hydrodynamics.
It was shown in Refs.\ \cite{lopez-fradkin-91,lopez-fradkin-98}, that
for systems with filling factors $\nu$ in the Jain
sequences~\cite{jain}, where $\displaystyle{\nu^{-1}=2n+p^{-1}}$ and $n$
and $p$ are integers, at a mean-field level the statistical field
screens the external magnetic field $B$ down to the effective field
$B_{\rm eff}=B/(2np+1)$. At this level of approximation the fermions
fill up an integer number $p$ of the Landau levels of $B_{\rm eff}$.
Naturally, this is just Jain's picture of composite fermions filling up
effective Landau levels~\cite{jain}. Fluctuations about this mean-field
state play a key role.

Now, if we integrate out the fermions, \textit{i.e.} we sum over
all current configurations including the effects of the fermion phase
factor as well as interaction effects, we will clearly obtain an
effective action for $\tilde a$ and $\tilde b$, the fluctuating pieces
of the statistical gauge field $a$ and the hydrodynamic gauge field $b$.
(Formally, this is done by expanding the fermion determinant in powers
of the gauge fields.):
\begin{equation}
S_{\rm fermion}=-i\hbar\tr\ln\left[ i\hbar D_0+\mu c-\frac{\hbar^2}{2Mc}\vec D^2\right]
+S_{\rm int}(\tilde b)
\end{equation}
where $S_{\rm int}(\tilde b)$ is the term of the action due to
electron-electron interactions; here we have used the constraint
relating the current to the hydrodynamic gauge field. The details of
this calculation are given explicitly by Lopez and Fradkin in Ref.\
\cite{lopez-fradkin-91,lopez-fradkin-93,lopez-fradkin-98}.
They computed the leading quadratic term in $\tilde a$, which amounts to
a calculation of the polarization tensor of electrons filling up $p$
effective Landau levels. The resulting expression is rather complex and
depends explicitly on the form of the interactions. However, to lowest
order in the momentum expansion, \textit{i.\ e.\/} at distances long
compared to the effective magnetic length $\ell_{\rm eff}$, it is simply
given by (see Fig.\ \ref{fig:2pt.eps})
\begin{equation}\label{eq:Seffta}
S_{\rm eff}[\tilde a]=\frac{\hbar p}{4\pi}\int d^3x\; 
\epsilon^{\mu \nu \lambda} \tilde a_\mu  \partial_\nu \tilde a_\lambda+\ldots
\end{equation}
\myfig{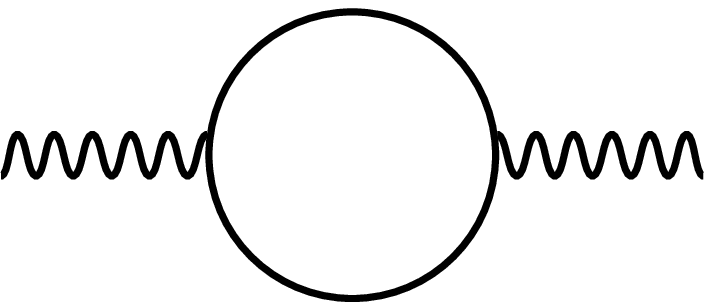}{4}{Two point function for $\tilde\csA$.}

Other effects, such as  the contributions from the electron-electron
interactions, appear at higher orders in derivatives (and in the parity
even sector only). Thus, to leading order we find the following
effective Lagrangian ~\cite{lopez-fradkin-98}
\begin{equation}
{\cal L}_{\rm eff}[\tilde a,\tilde b,\tilde A]= 
\frac{\hbar p}{4\pi} \epsilon^{\mu \nu \lambda} \tilde a_\mu \partial_\nu \tilde a_\lambda+
\frac{\hbar }{2\pi}   \epsilon^{\mu \nu \lambda} \tilde a_\mu \partial_\nu \tilde b_\lambda-
\frac{2n\hbar}{4\pi}  \epsilon^{\mu \nu \lambda} \tilde b_\mu \partial_\nu \tilde b_\lambda-
\frac{1}{2\pi}\frac{e}{c}  
\epsilon^{\mu \nu \lambda} \tilde A_\mu \partial_\nu \tilde b_\lambda
\label{eq:Zboth}
\end{equation}
where $\tilde A$ is a small fluctuation of the external electromagnetic
field. Note that if we integrate out $\tilde a$, at leading order we get
(up to boundary conditions)
\begin{equation}
{\cal L}_{\rm eff}[\tilde b]=-(2n+\frac{1}{p})\frac{\hbar}{4\pi}  
\epsilon^{\mu \nu \lambda} \tilde b_\mu \partial_\nu \tilde b_\lambda+\ldots
\end{equation}
The coefficient of ${\cal L}_{\rm eff}[\tilde b]$ is
$\displaystyle{\nu^{-1} = 2n+\frac{1}{p}}$. Thus $p=1$ corresponds to
the Laughlin series. In particular, also for $p=1$, this is the form of
the action that follows from hydrodynamic
arguments~\cite{wen-topo,wen-zee-matrix,wen-review,
frohlich-zee,frohlich-kerler}. It is straightforward to show
~\cite{lopez-fradkin-91} that this theory leads to the correct value of
the Hall conductivity
\begin{equation}
\sigma_{xy}=\frac{\nu}{2\pi} \;
\frac{e^2}{\hbar c}
\end{equation}
where
\begin{equation}
\frac{1}{\nu}=2n+\frac{1}{p}
\end{equation}
This result is exact because it saturates a Ward identity, {\it i.\
e.\/} the $f$-sum rule, as discussed in Refs.\
\cite{lopez-fradkin-91,zhang}.

The coefficient of the Chern-Simons action is known as the Chern-Simons
level. General topological consistency
arguments~\cite{witten-cs,hosotani} show that if the theory is quantized
on a closed manifold, the level of a Chern-Simons action must be
quantized for the theory to be invariant under large gauge
transformations. This requirement is met by the effective action of Eq.\
(\ref{eq:Zboth}), which contains both the statistical and the
hydrodynamic gauge fields, but not by either the effective action for
the statistical field or the hydrodynamic field alone except for the
Laughlin sequence, $p=1$. We will return to this issue in the context of
the non-commutative theory in Section \ref{sec:ncdual}.

\subsection{Trilinear Chern-Simons Couplings} \label{sec:res}
\newcommand{\mommeas}[1]{\frac{d^3#1}{2\pi\hbar}\ }

Hence, to quadratic order in fluctuations and at long distances, one
finds a commutative Abelian Chern-Simons gauge theory. We will
investigate now the trilinear terms in this expansion.

We have seen that to quadratic order in $\tilde a$, and to lowest order
in a gradient expansion, the effective action for a Laughlin FQH state
and its Abelian generalizations is a (commutative) Abelian Chern-Simons
theory. We will show now that the contributions to the effective action
trilinear in $\tilde a$ make the effective theory non-commutative. The
next correction, $S_{\rm eff}^{(3)}[\tilde a]$, is trilinear in the
fields  $\tilde a$, and it is shown in Fig. \ref{fig:3ptboth.eps}.
\myfig{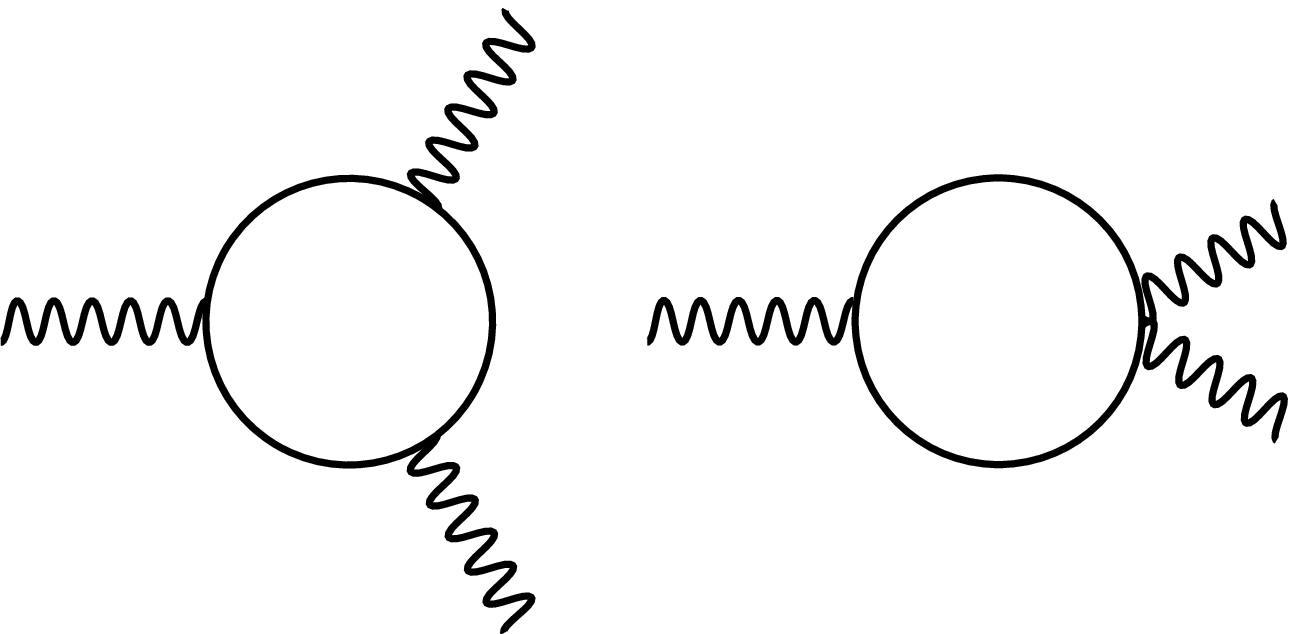}{5}{Diagrams contributing to the three point function
for $\tilde\csA$.} The details of this calculation are given in the Appendix. In momentum
space, we evaluate
\begin{equation}
S^{(3)}_{\rm eff}[\tilde a]=\frac{1}{3!}\int d^3P\ d^3Q\ d^3R\ 
\Pi^{\mu\nu\lambda}(P,Q,R)\tilde a_\mu(P)\tilde a_\nu(Q)\tilde a_\lambda(R)
\end{equation}
It is difficult to evaluate the resulting sums and integrals entering in
$\Pi^{\mu \nu \lambda}(P,Q,R)$ in closed form. They are given in the
form of a series with frequency poles at integer multiples of the
cyclotron frequency and with residues dictated by gauge invariance
multiplied by factors involving Laguerre polynomials of the variable
$P^2/2B_{\rm eff}$. The odd-parity terms of this tensor contain a phase
factor of the form
\begin{equation}
e^{\displaystyle{-i\ell_{\rm eff}^2 \; P\wedge Q}/2}
\label{eq:moyal}
\end{equation}
which we recognize as a Moyal phase. (To be more precise it is the
Fourier transform of Eq.\ (\ref{eq:Moyal}) with $\theta\to\ell_{\rm eff}^2$.) Notice that here $\ell_{\rm
eff}^2=\hbar c /e B_{\rm eff}$ is the effective magnetic length, which
is related to the magnetic length $\ell$ by
\begin{equation}
\ell_{\rm eff}^2=(2np+1) \ell^2=\frac{p}{\nu} \ell^2
\end{equation}
To leading order in an expansion in momenta there is one component
$\Pi^{012}$ which survives. We find, up to an energy-momentum conserving $\delta$-function
\begin{equation}\label{eq:pimoyal}
\Pi^{0[12]}(P,Q,R)=\frac{\hbar p}{2\pi} e^{-\displaystyle{i\ell_{\rm eff}^2 \; P\wedge Q}/2} 
\; \; f(\vec P^2,\vec Q^2)
\end{equation}
where $f(P^2,Q^2)$ is a regular function of the momenta $P$ and $Q$ and
$f(0,0)=1$.

We are being selective in keeping the momentum dependent phase. To be
fully consistent we should expand the exponential and group the momenta
into higher derivative corrections. However, we claim that this phase is
a Moyal phase indicative of the non-commutativity of this theory. This
phase automatically accompanies all Feynman diagrams, and at least at
one loop is the only odd parity contribution.

Thus, to the order of approximation we have kept, the effective
Lagrangian  for the statistical gauge fields $\tilde a$ is a
non-commutative Chern-Simons gauge theory, of the form of Eq.\
(\ref{eq:NCCS}), at level $p$:
\begin{equation}
{\cal L}_{\rm eff}[\tilde a]=\frac{\hbar p}{4\pi} \epsilon^{\mu \nu \lambda} 
\left(\tilde a_\mu \partial_\nu \tilde a_\lambda-
\frac{2i}{3} \tilde a_\mu \star \tilde a_\nu \star \tilde a_\nu \right) + \ldots
\label{eq:a-nc}
\end{equation}
where $\star$ denotes a Moyal product of Eq.\ (\ref{eq:moyal}) with
non-commutativity parameter $\theta_a=\ell_{\rm eff}^2$. Hence, to the
present order of approximation, the trilinear terms just calculated turn
the terms associated with  the statistical gauge field $\tilde a$ in the
effective action $S_{\rm eff}[a,b]$ of Eq.\ (\ref{eq:Zboth}) into a
non-commutative Chern-Simons gauge theory. The resulting effective
action still satisfies the quantization condition of the level of its
Chern-Simons terms.

However, there are a number of features of this result that seem to be
wrong. Although this effective theory  is non-commutative, the
non-commutativity is associated with the statistical gauge field $\tilde
a$ instead of the hydrodynamic gauge field $\tilde b$ as in Susskind's
action. In addition, the level of the non-commutative gauge theory is
$p$ instead of the denominator of the filling factor as suggested by
Ref.\ \cite{Suss}. Likewise, the length scale of the non-commutativity
is neither the average particle separation as in Susskind's action nor
the magnetic length $\ell$ as required by the magnetic current algebra,
but instead the effective magnetic length $\ell_{\rm eff}$. We will show
below, using the exact symmetries of the physical system, that at higher
orders there must be corrections which conspire to yield the correct
behavior for physical observables, such as the actual charge density and
current.

These corrections are not hard to find. In fact the expansion of the
fermion determinant leads to terms in the effective action for $\tilde
a$ with more than three external legs. In particular the terms with an
odd number of external legs have parity-odd pieces involving Moyal
products as well as more derivatives. These terms cannot be neglected in
the computation of the effective action for the hydrodynamic field
$\tilde b$.

\section{The non-commutative duality} \label{sec:ncdual}

Let us now make contact with Susskind's hydrodynamic effective theory.
We will do that by integrating out the statistical gauge field $a$, and
calculating an effective action for the hydrodynamic field $ b$. Since
the theory is now non-linear in $\tilde a$, we will do this in
perturbation theory. Thus, we go back and consider the full non-commutative
effective Lagrangian of Eq. (\ref{eq:Zboth}):
\begin{eqnarray}
{\cal L}_{\rm eff}[\tilde a,b]&=&
\frac{\hbar p}{4\pi} \epsilon^{\mu \nu \lambda}
\left\{ \tilde a_\mu \partial_\nu \tilde a_\lambda-
\frac{2i}{3}\tilde a_\mu \star \tilde a_\nu \star \tilde a_\lambda +\ldots\right\}+
\frac{\hbar}{2\pi}  \epsilon^{\mu \nu \lambda} \tilde a_\mu \partial_\nu b_\lambda-
\frac{2n\hbar}{4\pi} \epsilon^{\mu \nu \lambda}
b_\mu  \partial_\nu b_\lambda \nonumber \\
&&-
\frac{e}{2\pi c}  \epsilon^{\mu \nu \lambda} \tilde A_\mu \partial_\nu b_\lambda+\ldots
\nonumber 
\label{eq:Zbothnc}
\end{eqnarray}
and integrate out $\tilde a$ perturbatively. Here the ellipses denote terms
with more than three factors of $\tilde a$.

This expansion leads to an effective Lagrangian for the hydrodynamic
field $b$, the dual theory, of the form
\begin{equation}
{\cal L}_{\rm eff}[ b]=-\frac{\hbar}{4\pi \nu} \epsilon^{\mu \nu \lambda} \left( b_\mu
\partial_\nu  b_\lambda-\frac{2i}{3}  b_\mu \star
 b_\nu \star  b_\lambda+\ldots\right)-\frac{e}{2\pi c} \epsilon^{\mu \nu \lambda} \tilde
A_\mu \partial_\nu  b_\lambda.
\label{eq:dual}
\end{equation}
In order to write the effective action in this form, we have rescaled the $\star$-product. At the present level of
approximation, the effective non-commutativity parameter appearing in eq. (\ref{eq:dual}) is
\begin{equation}
\theta_b=\frac{1}{2np+1} \frac{\ell^2}{\nu} +\ldots
\label{eq:approximate-theta}
\end{equation} 
and the terms denoted by $\ldots$ are contributions to the trilinear
term in $ b$ originating from diagrams in the expansion on $\tilde a$ of
the type shown in Fig.\ \ref{fig:twoloop.eps}.
\myfig{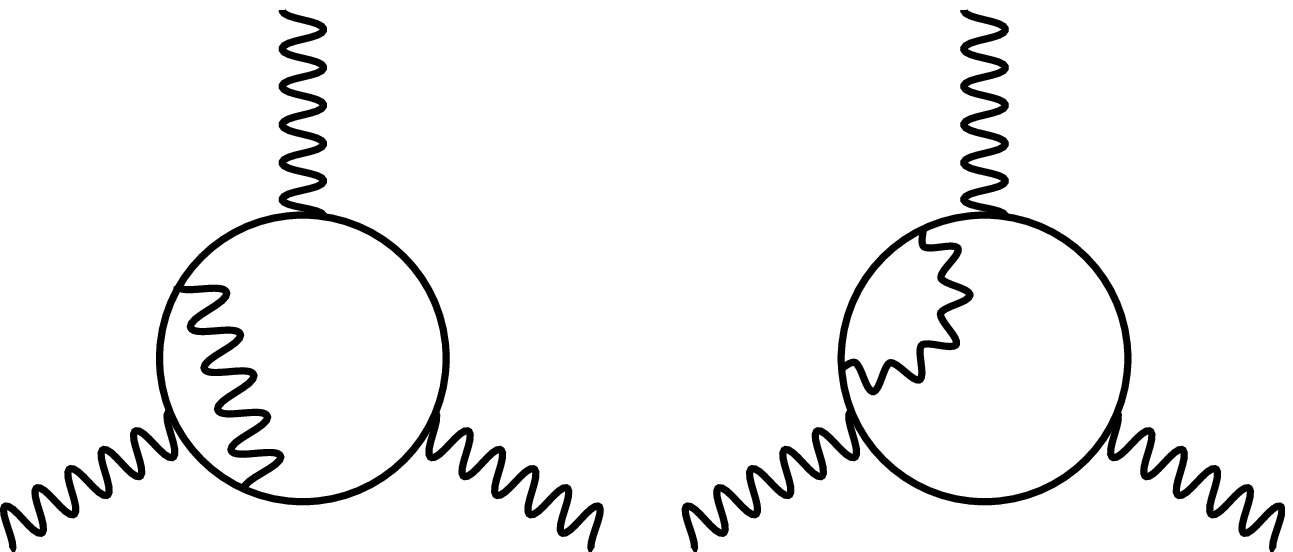}{5}{Two loop diagrams contributing to
$\Pi^{\mu\nu\lambda}$. The internal wavy lines are propagators of the
statistical gauge field $\tilde a$.} Such contributions, among other
things, lead to finite renormalizations of the non-commutativity
parameter.

For the particular case of  the Laughlin sequence where $p=1$ and
$\nu=1/(2n+1)$, we obtain an effective hydrodynamic theory of Susskind's
form at level $k=2n+1$ but with an effective non-commutativity parameter
$\theta=\ell^2/\nu k$. In contrast, in Susskind's theory~\cite{Suss} the
non-commutativity parameter is $\theta=\ell^2/\nu=1/(2\pi \rho)$,
\textit{i.\ e.\/} the length scale of non-commutativity is the mean
distance between electrons.

The effective hydrodynamic theory of Eq.\ (\ref{eq:dual}) is dual to the
theory described in terms of the statistical gauge field $a$. Under this
transformation particles and fluxes are interchanged. This is just the
particle-flux duality which underlies the physics of the fractional
quantum Hall states. Hence, the effective theory of  Eq.\
(\ref{eq:dual}) is the non-commutative extension of the duality picture
familiar from the commutative Chern-Simons theories of the fractional
quantum Hall effect~\cite{duality}.

The effective action $S_{\rm eff}[\tilde A]$ for an electromagnetic
perturbation $\tilde A$ can be computed by integrating out the
hydrodynamic field $b$. To lowest order we find
\begin{equation}
S_{\rm eff}=\frac{\nu e^2}{4\pi \hbar c^2} \int d^3x \; 
\epsilon^{\mu \nu \lambda} \left(\tilde A_\mu \partial_\nu \tilde A_\lambda-
 \frac{2i}{3}\; \frac{e}{\hbar c}  \tilde A_\mu \star \tilde A_\nu \star \tilde A_\lambda+\ldots \right)
\end{equation}
with a non-commutativity parameter which is now
\begin{equation}
 \theta_A=\displaystyle{\frac{\ell^2}{2np+1}+\ldots}
 \end{equation}
  This result implies that the Hall conductivity is $\sigma_{xy}=\nu
  e^2/h$, which is correct, but it predicts a non-commutativity
  parameter $\theta$ which is not equal to $\ell^2$. Below we will
  present an argument, essentially a Ward identity, which implies that
  for a translationally invariant system in a magnetic field the
  non-commutativity parameter should be exactly equal to $\ell^2$
  without any renormalization. This result also implies that the 
  non-commutativity parameter of the hydrodynamic theory must be exactly
  equal to $\ell^2/\nu=1/(2\pi \rho)$ as in Susskind's theory. Hence,
  the non-linear terms in the effective action for the statistical field
  $\tilde a$ must lead to perturbative contributions in the hydrodynamic
  theory whose total net effect is a finite renormalization of its
  non-commutativity parameter to the correct value $\ell^2/\nu$. In
  contrast the level of the Chern-Simons action cannot be renormalized
  as it is protected by topological invariance.

  \subsection{Magnetic Ward Identity} \label{sec:ward}

  In the previous sections we showed  that the Moyal phase appearing in 
  Eq.\ (\ref{eq:pimoyal}) indicates that the effective action for
  $\tilde\csA$ should be considered non-commutative. However this result
  led to a number of inconsistencies which we will discuss here.  As we
  noted above, the Moyal phase appearing in Eq.\  (\ref{eq:pimoyal}) is
  controlled by  $\Bf$, which sets the length scale relevant to the
  perturbation theory about the mean field state. However, as we noted
  above, $\Bf$ is not a physical scale.

  The mean field theory that we used here, usually called the average
  field approximation, is known to break a number of symmetries of the
  physical electron gas which must be preserved exactly. Let us
  reexamine what we have done. We first attached $2n$ flux quanta to
  each electron. This procedure does not change anything and it does not
  break any symmetry. However, at the mean field level, there are
  significant changes in the Hilbert space:  the system behaves like a
  system of composite fermions filling up an integer number $p$ of
  Landau levels of the  partially screened magnetic field $B_{\rm eff}$.
  In particular these effective levels are mixtures of the original
  Landau levels. This is the physics of the (unprojected) Jain wave
  functions~\cite{jain}. If taken literally, this mean field theory
  makes a number of wrong predictions. For instance it implies that the
  bosonic bound state with the leading residue, \textit{i.\ e.\/} a
  residue which vanishes like $\sim q^2$ as $\vec q\to 0$ as required by
  gauge invariance, has an energy gap at the effective cyclotron
  frequency $\omega_{\rm eff}=\omega/(2np+1)$. As it is well known, this
  result violates Kohn's theorem~\cite{kohn}, which states that in a
  translationally invariant system this gap must be at the exact
  cyclotron frequency $\omega_c=eB/mc$, without any mass
  renormalization. Kohn's theorem follows from the fact that the system
  as a whole must obey the magnetic algebra of the full magnetic field
  $B$. (A similar approximation also leads to a spurious zero-field Hall
  effect in theories of anyon superfluidity~\cite{cwwh}.) Similarly, the
  excitation spectrum predicted by this mean field theory is wrong,
  since it predicts that the excitations of this state are composite
  fermions instead of particles with fractional charge and fractional
  statistics. The problem with the length scale in the Moyal phase has
  the same origin.

  The solution  to these problems is well known. At the level of wave
  functions, the correct physical behavior is recovered  only after the
  states are projected onto the lowest Landau level~\cite{jain}. In the
  context of the Chern-Simons  theories, which deal with electrons as
  point particles and are not projected onto the lowest Landau level,
  the solution to these problems is also well understood. It has been
  shown in Refs.\ ~\cite{lopez-fradkin-91,lopez-fradkin-93,zhang}, that
  unlike other mean field theories, for fractional quantum Hall states,
  quantum fluctuations even at the Gaussian level, {\sl restore} some of
  the symmetries broken by the the mean field state. Physically this
  happens since, due to the Gauss Law constraint of the Chern-Simons
  theory, a local density fluctuation is equivalent to a local flux
  fluctuation. In particular, the leading quantum corrections, described
  by the Gaussian commutative effective action, yield upon integrating
  out the gauge fields $a$ and $b$, the correct electromagnetic
  polarization tensor which  saturates the $f$-sum rule in the $\vec q
  \to 0$ regime and thus it is exact in this
  limit~\cite{lopez-fradkin-91,lopez-fradkin-98,lopez-fradkin-93}.
  Hence, in the uniform limit this theory predicts, for the Jain states
  with filling factor $\nu=p/(2np+1)$, a Hall conductivity equal to
  $\sigma_{xy}=\frac{\nu}{2\pi} \frac{e^2}{\hbar}$, a cyclotron mode
  with frequency $\omega_c$ and residue $\sim\vec q^{\;2}$, a spectrum
  of quasiholes carrying charge $e/(2np+1)$ and fractional statistics
  $\pi/(2np+1)$, and, for a surface of genus $g$, a ground state
  degeneracy of $(2np+1)^g$. Thus the (Gaussian) commutative
  Chern-Simons effective action of Eq.\ (\ref{eq:Zboth}) yields a
  complete description of the universal long-distance data of fractional
  quantum Hall states (for a recent discussion see Ref.\ 
  \cite{lopez-fradkin-98}). (This is no longer the case when the gap
  collapses as in the limiting states $\nu=1/2n$, which are compressible
  states of the 2DEG~\cite{hlr}.)

  Let us see what this line of analysis implies for the non-commutative
  effective theory of Eq.\ (\ref{eq:Zboth}) and Eq.\ (\ref{eq:a-nc}). We
  noted before that in the $\vec q \to 0$ limit the Gaussian theory is
  exact. However, in this regime the theory is commutative. Although at
  shorter distances, the physics of the 2DEG is strongly non-universal,
  we showed above that in the odd-parity sector, the theory becomes
  non-commutative. However, the  length scale of the non-commutativity
  is $\ell_{\rm eff}$ which is not a physical length scale. This
  inconsistency has exactly the same origin as the violation of the
  global magnetic symmetry. As before, this inconsistency is  also
  resolved by quantum corrections to the mean field state.

  To understand this problem, let us look first at the actual
  electromagnetic currents $J_\mu$ in the system. (As usual, this is a
  $3$-vector formed by the physical charge density fluctuation and the
  two components of the charge current.) These currents are obtained as
  from the effective action for the external electromagnetic gauge field
  $S_{\rm eff}[\tilde A]=-\log Z[\tilde A_\mu]$, where $Z[\tilde A]$ is
  the partition function of the full 2DEG and $\tilde A$ is a weak
  electromagnetic perturbation. In 1992, Iso, Karabali and
  Sakita~\cite{sakita}, and independently Martinez and
  Stone~\cite{stone}, showed that the electromagnetic currents and
  charge densities for a 2DEG in a large magnetic field $B$, in an
  incompressible state with filling factor $\nu$, obey a $W_{\infty}$
  algebra, which is a local extension of the global magnetic algebra of
  finite translation operators in a magnetic field. Let $\hat \rho(x,y)$
  be the charge density operator {\sl projected on the Lowest Landau
  level}, and $\hat \rho(p,q)$ be its Fourier transform
\begin{equation}
\hat \rho(p,q)=\int dx dy \; \hat \rho(x,y)\; e^{\displaystyle{i(px+qy)/\hbar}}
\end{equation}
These authors~\cite{girvin-jach,sakita,stone} showed that the projected
density operators obey the algebra
\begin{equation}
\left[ \hat \rho(p,q),\hat \rho(p',q')\right]=-2i \sin\left(\ell^2(pq'-p'q)/2\right)\;
e^{\ell^2(pp'+qq')/2} \hat\rho(p+p',q+q')
\end{equation}
In the holomorphic basis 
\begin{equation}
\hat \rho_{m n}=\int dx dy \; \hat \rho(x,y)\; z^m \bar z^n 
\end{equation}
 the algebra becomes
\begin{equation}
\left[\hat \rho_{m n}, \hat \rho_{k l}\right]=\sum_{j=1}^{{\rm min} (n,k)} \frac{(-1)^j}{j!} 
\frac{n!k!}{(n-j)!(k-j)!} \hat \rho_{m+k-j,n+l-j}-(n \leftrightarrow l) (m \leftrightarrow k)
\end{equation}
Hence, the algebra of the charge density operators
~\cite{girvin-jach,sakita}, as well as of the current density operators
as shown in Ref.\ \cite{stone}, contains explicitly a Moyal product
controlled by the magnetic length scale $\ell$ of the full magnetic
field $B$. This is a very general result which follows from the
incompressibility of the system and from the properties of the quantum
states in a magnetic field.

This result poses important constraints, which can be regarded as Ward
identities, on the values of the parameters of the effective action
$S_{\rm eff}[\tilde A]$ of an external {\sl electromagnetic field} in an
incompressible quantum Hall state. In addition to the condition of gauge
invariance, which follows from current conservation, these results {\sl
require} that for a fractional quantum Hall state at filling factor
$\nu$, the effective action $S_{\rm eff}[\tilde A]$ must have the form
of a non-commutative Chern-Simons action whose level is $\nu$ (in units
of $e^2/\hbar$) and whose non-commutativity parameter
$\theta=\ell^2=\hbar c/eB$, where $B$ is the full magnetic field.  As we
saw above, this requirement in turn implies that the non-commutativity
scale for the hydrodynamic field is $1/(2\pi \rho)$, where $\rho$ is the
average areal density. Therefore, we conclude that the higher orders in
perturbation theory, among other things, must necessarily renormalize
the non-commutativity parameter of the hydrodynamic field to this exact
value.

\section{Conclusions} \label{sec:conclusions}

In this paper we have used the notion of flux attachment to derive
Susskind's non-commutative hydrodynamic action for the fractional
quantum Hall effect. We have presented two mutually dual effective
theories, which are non-commutative extensions of the familiar
(commutative) Chern-Simons descriptions of the FQHE. We have used this
duality to show that while the hydrodynamic effective action must have a
non-commutative parameter controlled by the average particle distance
(as in Susskind's action), the effective action of the electromagnetic
field must have a non-commutative parameter controlled only by the
magnetic length. We have also shown that while mean-field-like
approximations do lead to effective actions with the correct form of a
non-commutative Chern-Simons theory, the non-commutativity parameter is
not correct at lowest order and must acquire corrections in order to
satisfy the constraints imposed by the global magnetic algebra, and its
local (Moyal) extension. This is in marked contrast with the Hall
conductivity which is already correct at the level of the Gaussian
theory.

We would like to comment on the validity and usefulness of these
non-commutative effective theories. Clearly the commutative abelian
Chern-Simons theory gives a faithful description of the universal data
encoded in fractional quantum Hall fluids: the Hall conductivity, the
fractional charge and statistics of the excitations and the global
topological degeneracy of their ground states. The non-commutative
theory describes the local extension of the global magnetic algebra,
\textit{i.\ e.\/} the $W_\infty$ algebra of the local currents and
densities of incompressible charged two-dimensional fluids in large
magnetic fields. In fact, as argued by Susskind, non-commutative
Chern-Simons theory is a description of the area-preserving
diffeomorphisms of these incompressible fluids. Hence, in a sense,
except for the universal quantum numbers they encode, the information
contained in these effective theories is essentially kinematic in
nature. Indeed, there is much {\sl dynamics} taking place at the scale
of the magnetic (and Coulomb) lengths, such as the energetics of the
excitations,  which is not described by these effective theories.
Although in a formal sense this is an inconsistency, what matters here
is that the non-commutative terms included are the only contributions at
these length scales which are odd under parity and time-reversal.
Obviously a full description of the physics requires the (strongly
non-universal) microscopic physics which controls the energetics of
these states.

These considerations are particularly important for applications of
these ideas to non-uniform ground states such as Wigner crystals,
quantum Hall smectic and nematic states (for recent work see Refs.\
\cite{fogler,chalker,FK,bert,barci,fogler2,RD}). Recent work on these
interesting phases of the two-dimensional electron gas has revealed
that magnetic symmetry plays a crucial role in low-energy dynamics of
these states. Interestingly, recent work on non-commutative field
theories has suggested that their phase diagrams involve analogs of
these phases~\cite{gubser-sondhi}. It remains an important and open
problem to understand the implications and restrictions imposed by the
non-commutative structure on the phase transitions between these
states~\cite{sinova,moore}.

\section{Acknowledgments}

Discussions with Moshe Rozali and Michael Stone are gratefully
acknowledged. Work supported in part by U.\ S.\ Department of Energy,
grant DE-FG02-91ER40677 (RL), and by the National Science Foundation
through the grant DMR-01-32990 (EF).  VJ has been supported by a GAANN
fellowship from the U.S. Department of Education, grant 1533616.

\appendix
\section{}
\label{sec:appendix}


In this appendix, we give details of the perturbative computations. For a detailed discussion of the physics behind this theory, see Ref. \cite {lopez-fradkin-91}. We borrow their notation, and pick up the discussion at an appropriate point. Within this appendix, we have set $e=\hbar=c=1$, and thus $\Bf=1/\ell_{\rm eff}^2$. The Feynman rules are obtained from the
Lagrangian density for non-relativistic fermions in a magnetic field:
\begin{equation}
{\cal L}=\psi^\dagger
L\psi-\csA_0\psi^\dagger\psi+\frac{i}{2M}\csA_j\left[(D_j^\dagger\psi^\dagger)\psi-
\psi^\dagger D_j\psi\right]+\frac{1}{2M}\vec\csA^2\psi^\dagger\psi
\end{equation}
where 
\begin{eqnarray}
D=\pa+iA,\ \ \ \ \ D^\dagger=\pa-iA\\
L=i\pa_0-A_0+\mu+(1/2m)\vec D^2
\end{eqnarray}
The propagator is written
$\langle\psi(x_1)\psi^\dagger(x_2)\rangle=iG(x_1,x_2).$
It is convenient for this calculation to rescale spatial coordinates $\vec u_i=\sqrt{\Bf}\ \vec x_i$ and similarly rescale
momenta $\vec P_i\to \vec P_i/\sqrt{\Bf}$. We then have
\begin{equation}
iG(u,v)=\int \frac{dk}{2\pi}\ S_{x^0,y^0,m}
\varphi_{m,k}(\vec u)\varphi^*_{m,k}(\vec v)
\end{equation}
where for brevity we have written the summation symbol
\begin{equation}
S_{x^0,y^0,m}=\left(\theta(x^0-y^0)\sum_{m=p}^\infty-
\theta(y^0-x^0)\sum_{m=0}^{p-1}\right) e^{-i\omega_m(x^0-y^0)}
\end{equation}
and
\begin{equation}
\varphi_{m,k}(\vec u)={\cal N}_m e^{ik u_{x}}\ e^{-(u_{y}-k)^2/2}\ H_m(u_{y}-k)
\end{equation}
with normalization ${\cal N}_m=(2^{m}m!\sqrt{\pi})^{-1/2}$.
We choose a gauge such 
that 
\begin{eqnarray}
\frac{iD^{(x)}_x}{M}&=&\alpha\left(i\frac{\pa}{\pa u_x}+u_y\right)\equiv d_x^{(u)}\\
\frac{iD^{(x)}_y}{M}&=&i\alpha \frac{\pa}{\pa u_y}\equiv d_y^{(u)}.
\end{eqnarray}
where $\alpha=\sqrt{\Bf}/M$. These derivatives have the effect of shifting the indices of the Landau functions:
\begin{eqnarray}
d_a^{(u_1)} iG(u_1,u_2)=\alpha\int \frac{dk}{2\pi}\ 
S_{x^0,y^0,m}
\sum_Jc_{m,a}^{(J)}\varphi_{m+J,k}(\vec u_1)\varphi^*_{m,k}(\vec u_2)\nonumber\\
d_a^{(u_1)\dagger} iG(u_3,u_1)=\alpha\int \frac{dk}{2\pi}\ S_{x^0,y^0,m}
\varphi_{m,k}(\vec u_3)\sum_{\tilde J}\tilde c_{m,a}^{(\tilde J)}
\varphi^*_{m+\tilde J,k}(\vec u_1)\nonumber
\end{eqnarray}
where we have introduced constants $c$ and $\tilde c$, whose non-zero components are:
\begin{eqnarray}
c_{m,x}^{(1)}&=\sqrt{\frac{m+1}{2}},\ \ \ \ \ \ \ c_{m,y}^{(1)}=-i\sqrt{\frac{m+1}{2}}\\
c_{m,x}^{(-1)}&=\sqrt{\frac{m}{2}},\ \ \ \ \ \ \ c_{m,y}^{(-1)}=+i\sqrt{\frac{m}{2}}\\
\tilde c_{m,x}^{(1)}&=\sqrt{\frac{m+1}{2}},\ \ \ \ \ \ \ \tilde c_{m,y}^{(1)}=
+i\sqrt{\frac{m+1}{2}}\\
\tilde c_{m,x}^{(-1)}&=\sqrt{\frac{m}{2}},\ \ \ \ \ \ \ \tilde c_{m,y}^{(-1)}=
-i\sqrt{\frac{m}{2}}
\end{eqnarray}
In the three-point Feynman diagrams (Figs. \ref{fig:3ptboth.eps}), for each
spatial component of a gauge field that appears at an external leg, there is
a spatial derivative of the form given above acting on the fermion lines. If
we sum over all such possibilities, the net effect on a vertex is to replace
its Landau functions\footnote{The Landau functions come from the fermion
propagator, but at each vertex there are two such functions.} by a factor
\begin{equation}
\sum_{\tilde J_1,J_2}C_{\tilde J_1,J_2}(m_1,m_2;\mu)\varphi^*_{m_1+\tilde J_1,k_1}
(\vec u_2)\varphi_{m_2+J_2,k_2}(\vec u_2).
\end{equation}
It is not difficult to show that
\begin{eqnarray}
C_{\tilde J_1,J_2}(m_1,m_2;0)&=&\delta_{\tilde J_1,0}\delta_{J_2,0}\\
C_{\tilde J_1,J_2}(m_1,m_2;a)&=&\frac{\alpha}{2}\left( c_{m_2,a}^{(J_2)}\delta_{\tilde J_1,0}+
\tilde c_{m_1,a}^{(\tilde J_1)}\delta_{J_2,0}\right),\ \ \ \ \ a=x,y.
\end{eqnarray}
$C$ is essentially the ``Fourier transformed'' three point vertex of $\psi^\dagger a_\mu \psi$. 

We write the correction to the effective action as
\begin{equation}
\delta S_{eff}=\frac{1}{3!}\int d^3x_1 d^3x_2 d^3x_3\ a_{\mu_1} (x_1) a_{\mu_2} (x_2) a_{\mu_3} (x_3) \Pi^{\mu_1\mu_2\mu_3}(x_1,x_2,x_3)
\end{equation}
where $\Pi$ is
\begin{eqnarray}
\int \sum_{\vec J,\vec{\tilde J}}\prod_{j=1}^3\left[
\frac{dk_j}{2\pi}S_{x_j^0,x_{j+1}^0,m_j}C_{\tilde J_j,J_{j+1}}(m_j,m_{j+1};\mu_{j+1})
\varphi_{m_j+J_j,k_j}(\vec u_j)\varphi^*_{m_{j-1}+\tilde J_{j-1},k_{j-1}}(\vec u_j)\right]
\end{eqnarray}
(Note: actually, this is just one of the two Feynman diagrams. The other can contribute only to $\Pi^{aa\mu}$.)
To proceed, we Fourier transform the CS fields
\begin{equation}
a_\mu (x)=\int dP^0 d^2P a_\mu(P) e^{iP^0x^0}e^{-i\vec P\cdot\vec u}
\end{equation}
Note that $\vec P$ here is the rescaled momentum.

It is then a simple matter to perform the integrals over the time coordinates; the net
effect is that the sums and $\theta$-functions are replaced by
%
$2\pi\delta(P_0+Q_0+R_0)$ times
\begin{eqnarray}
T=\frac{\left[\sum_{m_1=p}^\infty\sum_{m_2=p}^\infty\sum_{m_3=0}^{p-1}-\sum_{m_1=0}^{p-1}\sum_{m_2=0}^{p-1}\sum_{m_3=p}^\infty
\right]}{
(\omega_{m_3}-\omega_{m_1}+P_0)(\omega_{m_3}-\omega_{m_2}-R_0)}
+{\rm cyclic}
\end{eqnarray}
where we cycle simultaneously on $P_0,Q_0,R_0$ and $m_1,m_2,m_3$.


Next, we can do the spatial integrals, using the 
orthogonality for the $\varphi$'s
\begin{equation}
\int d^2u\ \varphi_{m,k}(\vec u)\varphi^*_{m',k'}(\vec u) e^{-i\vec P\cdot\vec u}
=2\pi\delta(k-k'-P_x)e^{-\vec P^2/4}e^{-iP_y(k+k')/2}
{\cal L}_{m,m'}(\vec P)
\end{equation}
where
\begin{eqnarray}
{\cal L}_{m,m'}(\vec P)&=&
\left( \theta(m-m')\sqrt{\frac{m'!}{m!}}\left(-{\cal P}\right)^{m-m'}
L_{m'}^{m-m'}(|{\cal P}|^2)\right.\\
&&+\left.\theta(m'-m)\sqrt{\frac{m!}{m'!}}\left({\cal P}^*\right)^{m'-m}
L_{m}^{m'-m}(|{\cal P}|^2)\right)
\end{eqnarray}
where we have defined ${\cal P}=(P_x+iP_y)/\sqrt{2}$.

Thus we find
\begin{eqnarray}
\int\prod_{j=1}^{3}\left[\frac{dk_j}{2\pi}\
2\pi\delta(k_j-k_{j-1}-P_{j,x})
e^{-\vec P_j^2/4}
e^{-iP_{j,y}(k_j+k_{j-1})/2}\right]\times\nonumber \\
T\left\{\sum_{\vec J,\vec{\tilde J}} \prod_{j=1}^3
C_{\tilde J_{j-1},J_j}(m_{j-1},m_j;\mu_j){\cal L}_{m_j+J_j,m_{j-1}+\tilde J_{j-1}}(\vec P_j)\right\}
\end{eqnarray}
Performing the $k_j$-integrations, we find
\begin{eqnarray}
\frac{1}{2\pi}e^{-i(P_{1,x}P_{2,y}-P_{1,y}P_{2,x})/2}
T\left\{\sum_{\vec J,\vec{\tilde J}} \prod_{j=1}^3e^{-\vec P^2_j/4}
C_{\tilde J_{j-1},J_j}(m_{j-1},m_j;\mu_j){\cal L}_{m_j+J_j,m_{j-1}+\tilde J_{j-1}}(\vec P_j)\right\}
\label{eq:result}
\end{eqnarray}
(We have dropped an overall
energy-momentum conserving $\delta$-function.)
We note that this result carries an overall Moyal phase, times a complicated function
of energies and momenta.

To proceed further, we need to evaluate the rather complicated sums in eq. (\ref{eq:result}). We will do so by expanding in energies and momenta (keeping the overall Moyal phase intact). We find that there is a non-zero contribution to 
the parity odd polarization
\begin{equation}\label{eq:pimoyal2}
\Pi^{0[12]}(P_1,P_2,P_3)=\frac{ p}{2\pi} e^{\displaystyle{-i\ell_{\rm eff}^2 \; P_1\wedge P_2/2}} 
\; \; f( P_1^2,P_2^2)
\end{equation}
where $f(P^2,Q^2)$ is a regular function of the momenta $P$ and $Q$ and $f(0,0)=1$.
Contributions to other polarizations of $\Pi$ are also calculable, but are of no relevance to the discussions of this paper.


\end{document}